\documentclass[12pt]{article}
\usepackage{a4wide}
\usepackage{epsfig}
\usepackage{amsmath}

\newlength{\absize}
\catcode`@=11
\def\citer{\@ifnextchar [{\@tempswatrue\@citexr}{\@tempswafalse\@citexr[]}}

\def\@citexr[#1]#2{\if@filesw\immediate
  \write\@auxout{\string\citation{#2}}\fi
  \def\@citea{}\@cite{\@for\@citeb:=#2\do
    {\@citea\def\@citea{--\penalty\@m}\@ifundefined
       {b@\@citeb}{{\bf ?}\@warning
       {Citation `\@citeb' on page \thepage \space undefined}}%
\hbox{\csname b@\@citeb\endcsname}}}{#1}}
\catcode`@=12

\begin{document}
  \thispagestyle{empty}
  \pagestyle{empty}
  \renewcommand{\thefootnote}{\fnsymbol{footnote}}
\newpage\normalsize
    \pagestyle{plain}
    \setlength{\baselineskip}{4ex}\par
    \setcounter{footnote}{0}
    \renewcommand{\thefootnote}{\arabic{footnote}}
\newcommand{\preprint}[1]{%
  \begin{flushright}
    \setlength{\baselineskip}{3ex} #1
  \end{flushright}}
\renewcommand{\title}[1]{%
  \begin{center}
    \LARGE #1
  \end{center}\par}
\renewcommand{\author}[1]{%
  \vspace{2ex}
  {\Large
   \begin{center}
     \setlength{\baselineskip}{3ex} #1 \par
   \end{center}}}
\renewcommand{\thanks}[1]{\footnote{#1}}
\vskip 0.5cm

\begin{center}
{\large \bf Test of Quantum Effects of Spatial Noncommutativity
using Modified Electron Momentum Spectroscopy}
\end{center}
\vspace{1cm}
\begin{center}
Jian-Zu Zhang$\;^{1, {\ast}}$, Ke-Lin Gao$\;^2$, Chuan-Gang
Ning$\;^3$
\end{center}
\begin{center}
$^1$ Institute for Theoretical Physics, East China University of
Science and Technology, Box 316, Shanghai 200237, P. R. China \\
$^2$ State Key Laboratory of Magnetic Resonance and Atomic and
Molecular Physics, Wuhan Institute of Physics and Mathematics,
Chinese Academy of Science, Wuhan 430071, P. R. China\\
$^3$ Department of Physics and Key Laboratory of Atomic and
Molecular NanoSciences of MOE, Tsinghua University, Beijing
100084, P. R. China
\end{center}
\vspace{1cm}

\begin{abstract}
The possibility of testing spatial noncommutativity by current
experiments on normal quantum scales is investigated.
For the case of both position-position and momentum-momentum
noncommuting spectra of ions in crossed electric and magnetic
fields are studied in the formalism of noncommutative quantum
mechanics.
In a limit of the kinetic energy approaching its lowest eigenvalue
this system possesses non-trivial dynamics.
Signals of spatial noncommutativity in the angular momentum are
revealed. They are within limits of the measurable accuracy of
current experiments.
An experimental test of the predictions using a modified electron
momentum spectrum is suggested. The related experimental
sensitivity and subtle points are discussed.
The results are the first step on a realizable way towards a
conclusive test of spatial noncommutativity.
\end{abstract}

\begin{flushleft}
$^{\ast}$ Corresponding author
\end{flushleft}
\clearpage

Quantum theory in noncommutative space \citer{CDS,MMP}
presents an
attractive possibility as a candidate in the present round in
hinting at new physics. In the low energy aspect, quantum
mechanics in noncommutative space (NCQM) \citer{CST,BRACZ},
\cite{JZZ04a} have been studied in detail. But testing their
predictions require experiments near the Planck scale which are
un-realizable, and /or their modifications to normal quantum
theory depending on vanishingly small noncommutative parameters
which are outside limits of measurable accuracy of current
experiments. It seems that noncommutative quantum theory escapes
measurement on normal quantum scales.

The possibility of testing spatial noncommutativity by current
experiments on normal quantum scales was investigated, and two
proposals \cite{JZZ04a} using Rydberg atoms
and Chern-Simons processes
are suggested. The two proposals revealed that in a limit of
diminishing the magnetic field to zero the vanishingly small
noncommutative parameters usually present in predictions derived
from spatial noncommutativity actually cancel out in the the
angular momentum, so that the lowest angular momentum turns out to
be $\hbar/4$. This provides a conclusive test of spatial
noncommutativity, i. e., a positive experimental result shows an
evidence of spatial noncommutativity and a negative one draws a
conclusive preclusion of spatial noncommutativity.
In practice the magnetic field, however, can only reach some
finite value limited by the level of shielding the background
magnetic fields. In order to meet the condition of a cancellation
between the vanishingly small noncommutative parameters present in
the angular momentum derived from spatial noncommutativity, the
magnetic field must be decreased to a level of some orders less
than the effective intrinsic magnetic field $B_{\eta}$ originated
from spatial noncommutativity (see below).
%
The field control at that level seems close to impossible in the
foreseeable future.

According to the present level of shielding the background
magnetic fields, this paper explores a realizable way for testing
noncommutative quantum effects on normal quantum scale.
For the case of both position-position and momentum-momentum
noncommuting spectra of ions trapped in crossed electric and
magnetic fields are investigated.
In a limit of the kinetic energy approaching its lowest eigenvalue
this system possesses non-trivial dynamics. The corresponding
constraints are analyzed.
Signals of spatial noncommutativity in the angular momentum are
revealed. They are within limits of the measurable accuracy of
current experiments.
An experimental test of the predictions using, similar to electron
momentum spectroscopy (EMS) \cite {MW}, a modified EMS is
suggested, and the related experimental sensitivity and the subtle
points are discussed.
It is the first step on a realizable way towards a conclusive test
of spatial noncommutativity.

We consider an ion of mass $\mu$ and charge $q(>0)$ trapped in a
uniform magnetic field ${\bf B}$ aligned along the $x_3$-axis and
an electrostatic potential \citer{trap}
\begin{equation}
\label{Eq:V}
V_{eff}=
\frac{1}{2}\mu\left[\omega_{\rho}^2(x_1^2+x_2^2)+\omega_z^2x_3^2\right]
\end{equation}
where
$\omega_{\rho}^2$ and
$\omega_z^2$
are
frequencies, respectively, in the $(x_1,x_2)$-plane
and $z$ direction.
The vector potential $A_i$ of ${\bf B}$ is chosen as
$A_i=-B\epsilon_{ij}x_j/2$, $A_3=0,\; (i,j=1,2).$
The Hamiltonian $H$ of the trapped ion can be decomposed into a
one-dimensional harmonic Hamiltonian $H_z$ in the $z$ direction
and a two-dimensional Hamiltonian $H_2$ in the $(x_1,x_2)$-plane:
$H=\left(p_i-qA_i/c\right)^2/2\mu+V_{eff}=H_2+H_z,$
%
$H_z=p_3^2 /2\mu+\mu\omega_z^2 x_3^2/2,$
and $H_2$ is (Henceforth the summation
convention is used):
\begin{equation}
\label{Eq:H2}
H_2=\frac{1}{2\mu}\left(p_i+
\frac{1}{2}\mu\omega_c\epsilon_{ij}x_j\right)^2+
\frac{1}{2}\kappa x_i^2=
\frac{1}{2\mu}p_i^2+\frac{1}{2}\omega_c\epsilon_{ij} p_i
x_j+\frac{1}{2}\mu\omega_P^2 x_i^2,
\end{equation}
where $\kappa=\mu\omega_{\rho}^2$, $\omega_c=qB/\mu c$ (the
cyclotron frequency), and
$\omega_P=(\omega_{\rho}^2+\omega_c^2/4)^{1/2}.$

If NCQM is a realistic physics, low energy quantum phenomena
should be reformulated in the formalism of NCQM. We consider the
case of both position-position noncommutativity (position-time
noncommutativity is not considered) and momentum-momentum
noncommutativity. The consistent deformed Heisenberg-Weyl algebra
\cite{JZZ04a} is:
$$[\hat x_{I},\hat x_{J}]=i\xi^2\theta_{IJ}, \qquad [\hat
x_{I},\hat p_{J}]=i\hbar\delta_{IJ}, \qquad [\hat p_{I},\hat
p_{J}]=i\xi^2\eta_{IJ},\;(I,J=1,2,3)$$
where $\theta_{IJ}$ and $\eta_{IJ}$ are the antisymmetric constant
parameters, independent of the position and momentum. We define
$\theta_{IJ}=\epsilon_{IJK}\theta_K,$ where $\epsilon_{IJK}$ is a
three-dimensional antisymmetric unit tensor. We put
$\theta_3=\theta$ and the rest of the $\theta$-components to zero
(which can be done by a rotation of coordinates), then we have
$\theta_{ij}=\epsilon_{ij}\theta$ $(i,j=1,2)$, where
$\epsilon_{ij}=\epsilon_{ij3}$ is a two-dimensional antisymmetric
unit tensor with $\epsilon_{12}=-\epsilon_{21}=1,$
$\epsilon_{11}=\epsilon_{22}=0$. Similarly, we have
$\eta_{ij}=\epsilon_{ij}\eta$. Thus we obtain the following two
dimensional deformed Heisenberg-Weyl algebra
\begin{equation}
\label{Eq:xp}
[\hat x_i,\hat x_j]=i\xi^2\theta_{ij}, \qquad [\hat x_i,\hat
p_j]=i\hbar\delta_{ij}, \qquad [\hat p_i,\hat
p_j]=i\xi^2\eta_{ij}.\;(i,j=1,2)
\end{equation}
Here we consider the noncommutativity of the intrinsic canonical
momentum. It means that the parameter $\eta$, like the parameter
$\theta$, should be extremely small.
This is guaranteed by a direct proportionality between them (See
Eq.~(\ref{Eq:eta-theta}) below). In Eqs.~(\ref{Eq:xp}) the scaling
factor $\xi$ is
$\xi=(1+\theta\eta/4\hbar^2)^{-1/2}.$

The deformed Heisenberg - Weyl algebra (\ref{Eq:xp}) can be
realized by undeformed variables $x_{i}$ and $p_{i}$ as follows
\begin{equation}
\label{Eq:hat-x-p}
\hat x_{i}=\xi(x_{i}-\frac{1}{2\hbar}\theta\epsilon_{ij}p_{j}),
\quad
\hat p_{i}=\xi(p_{i}+\frac{1}{2\hbar}\eta\epsilon_{ij}x_{j}),
\end{equation}
where $x_{i}$ and $p_{i}$ satisfy the undeformed Heisenberg-Weyl
algebra
$[x_{i},x_{j}]=[p_{i},p_{j}]=0,\;
[x_{i},p_{j}]=i\hbar\delta_{ij}.$
It should be emphasized that for the case of both
position-position and momentum-momentum noncommuting the scaling
factor $\xi$ in Eqs.~(\ref{Eq:xp}) and (\ref{Eq:hat-x-p})
guarantees consistency of the framework, and plays an essential
role in dynamics.
%
%

The deformed Hamiltonians $\hat H_2$ and $\hat H_z$ in
noncommutative space can be obtained by reformulating the
corresponding undeformed ones in commutative space in terms of
deformed canonical variables $\hat x_i$ and $\hat p_i$. Because of
$\hat x_3=x_3$ and $\hat p_3=p_3$ the deformed $\hat H_z(\hat
x_3,\hat p_3)$ is the same as the undeformed one, $\hat H_z(\hat
x_3,\hat p_3)=H_z(x_3,p_3)$.

The deformed $\hat H_2(\hat x,\hat p)$, using
Eqs.~(\ref{Eq:hat-x-p}), can be further represented by undeformed
variables $x_i$ and $p_i$ as
\begin{eqnarray}
\label{Eq:CS-H1}
\hat H_2(\hat x,\hat p)=\frac{1}{2M}(p_i+\frac{1}{2}G\epsilon_{ij}
x_j)^2 +\frac{1}{2}K x_i^2 =\frac{1}{2M}
p_i^2+\frac{1}{2M}G\epsilon_{ij} p_i x_j+\frac{1}{2}M\Omega_P^2
x_i^2,
\end{eqnarray}
where the effective parameters 
$M, G, \Omega_P$ and $K$ are defined as
\begin{eqnarray}
\label{Eq:M-K}
1/2M&\equiv&\xi^2(c_1^2/2\mu+\kappa\theta^{\;2}/8\hbar^2),\;
G/2M\equiv\xi^2(c_1 c_2/\mu+\kappa\theta/2\hbar),
\nonumber\\
M\Omega_P^2&\equiv&\xi^2(c_2^2/\mu+\kappa),\quad\quad\quad\quad
K\equiv M\Omega_P^2-G^2/4M,
\end{eqnarray}
and
$c_1=1+\mu\omega_c\theta/4\hbar,\;c_2=\mu\omega_c/2+\eta/2\hbar.$

The deformed Hamiltonian (\ref{Eq:H2}) and the equivalent one of
Eq.~(\ref{Eq:CS-H1}) possess a rotational symmetry in $(x_1,
x_2)$-plane. The $z$-component of the orbital angular momentum is
a conserved observable.

In order to explore new features of such a system we need to
investigate a Chern-Simons term
$\hat J_z=\epsilon_{ij}\hat x_i\hat p_j.$
From the NCQM algebra (\ref{Eq:xp}) we obtain commutation
relations between $\hat J_z$ and $\hat x_i,$ $\hat p_i$:
$[\hat J_z,\hat x_i]=i\epsilon_{ij}\hat x_j+i\xi^{2}\theta \;\hat
p_i, \quad
[\hat J_z,\hat p_i]=i\epsilon_{ij}\hat p_j-i\xi^{2} \eta\;\hat
x_i.$
From the above commutation relations we conclude that $\hat J_z$
plays approximately the role of the generator of rotations at the
deformed level.
Using Eqs.~(\ref{Eq:hat-x-p}), we can further represent
$\hat J_z$ by undeformed variables $x_i$ and $p_i$ as
\begin{equation}
\label{Eq:J1}
\hat J_z=\epsilon_{ij}x_ip_j-\frac{1}{2\hbar}\xi^{2}\left(\theta
p_i p_i +\eta x_i
x_i\right)=J_z-\frac{1}{2\hbar}\xi^{2}\left(\theta p_i p_i +\eta
x_i x_i\right),
\end{equation}
where $J_z=\epsilon_{ij}x_ip_j$ is the $z$-component of the
orbital angular momentum in commutative space. $\hat J_z$ and
$\hat H_2$ commute each other. They have common eigenstates.


We investigate an interesting case: dynamics of this system in the
limit of the mechanical kinetic energy approaching its lowest
eigenvalue.
For discussing this limit it is convenient to work in the
Lagrangian formalism. The Lagrangian corresponding to the
Hamiltonian $\hat H_2$ in Eq.~(\ref{Eq:CS-H1}) is
\begin{equation}
\label{Eq:L1}
\hat L = \frac{1}{2}M\dot{x_i}\dot{x_i}+
\frac{1}{2}\epsilon_{ij}\dot{x_i}x_j- \frac{1}{2}K x_i x_i.
\end{equation}
The mechanical kinetic energy
$\hat H_k=M\dot{x_i} \dot{x_i}/2$
can be rewritten as
\begin{equation}
\label{Eq:Hk2}
\hat H_k= \frac{1}{2M}\left(K_1^2+K_2^2\right)
\end{equation}
where
\begin{equation}
\label{Eq:K1}
K_i\equiv p_i+\frac{1}{2}G\epsilon_{ij}x_j
\end{equation}
are the mechanical momentum corresponding to the vector potentials
$A_i$. They satisfy the commutation relation
\begin{equation}
\label{Eq:K-K}
[K_i,K_j]=i\hbar G\epsilon_{ij}.
\end{equation}
In Eq.~(\ref{Eq:K1}) $p_i=\partial/\partial x_i$ are the canonical
momentum. They satisfy the commutation relation $[p_i,p_j]=0.$

We define canonical variables $Q=K_1/G$ and $\Pi=K_2,$ which
satisfy
\begin{equation}
\label{Eq:Q-Pi}
[Q,\Pi]=i\hbar\delta_{ij}.
\end{equation}
The kinetic energy $\hat H_k$ in Eq.~(\ref{Eq:Hk2}) is rewritten
as a Hamiltonian of a harmonic oscillator
\begin{equation}
\label{Eq:Hk3}
\hat H_k=\frac{1}{2M}\Pi^2+\frac{1}{2}M\omega_0^2 Q^2,
\end{equation}
where the effective frequency $\omega_0\equiv G/M.$ The
eigenvalues of $\hat H_k$ are
\begin{equation}
\label{Eq:Ekn}
\mathcal{\hat
E}_{k,n}=\hbar\omega_0\left(n+\frac{1}{2}\right),\;(n=0,1,2,\cdots).
\end{equation}
Its lowest one is
\begin{equation}
\label{Eq:Ek0}
\mathcal{\hat E}_{k,0}=\frac{\hbar G}{2M}.
\end{equation}
In the limit of $\hat H_k \to \mathcal{\hat E}_{k,0}$ the
Hamiltonian $\hat H_2$ in Eq.~(\ref{Eq:CS-H1})
reduces to
\begin{equation}
\label{Eq:H0}
\hat H_0=\frac{1}{2}K x_i x_i+\mathcal{\hat E}_{k,0}.
\end{equation}
The Lagrangian corresponding to $\hat H_0$ is
\begin{equation}
\label{Eq:L0}
\hat L_0=\frac{1}{2}G\epsilon_{ij}\dot{x_i}x_j-
\frac{1}{2}K x_i x_i-\mathcal{\hat E}_{k,0}.
\end{equation}

In the following we demonstrate that the reduced system $(\hat
H_0,\hat L_0)$ has non-trivial dynamics. The canonical momenta
\begin{equation}
\label{Eq:pi}
p_i=\frac{\partial\hat
L_0}{\partial\dot{x_i}}=\frac{1}{2}G\epsilon_{ij}x_j.
\end{equation}
The  Hamiltonian $\hat H_0^{\prime}$ obtained from $\hat L_0$ is
$\hat H_0^{\prime}=p_i\dot{x_i}-\hat L_0=K x_i x_i/2+\mathcal{\hat
E}_{k,0},$
which is just $\hat H_0$.

The canonical momenta $p_i$ in Eq.~(\ref{Eq:pi}) does not
determine velocities $\dot{x_i}$ as functions of $p$ and $x$ which
indicates that $\hat L_0$ is singular, but gives relations among
$p$ and $x$. Such relations are primary constraints
\cite{Baxt,JZZ96}
\begin{equation}
\label{Eq:C1}
\varphi_i(x,p)=p_i+\frac{1}{2}G\epsilon_{ij} x_j=0,
\end{equation}
The physical meaning of Eq.~(\ref{Eq:C1}) is that it expresses the
dependence of degrees of freedom among $p$ and $x$. The
constraints (\ref{Eq:C1}) should be carefully treated.

The Hamiltonian equation of $\hat H_0$ in Eq.~(\ref{Eq:H0})
gives
$\dot{x_i}=\partial \hat H_0/\partial p_i=0.$
But the $\hat L_0$ in Eq.~(\ref{Eq:L0}) has non-vanishing
$\dot{x_i}$. This needs to be clarified. The Hamiltonian equations
of such a singular (constrained) system are not unique. Because of
the constraints $\varphi_i(x,p)=0$ of Eq.~(\ref{Eq:C1}), $\hat
H_0$ plus any linear combination of $\varphi_i$ is also a
Hamiltonian of the system, i. e., the $\hat H_0$ can be replaced
by
$\hat H_0(x,p)+\lambda_i\varphi_i(x,p)$
where the Lagrange multiplier $\lambda_i$ may be a function of $x$
and $p$. The Hamiltonian equations, including the contributions of
$\delta \left(\lambda_i(x,p)\varphi_i(x,p)\right),$
read
\begin{equation}
\label{Eq:Heq}
\dot{p_i}=-\frac{\partial \hat H_0}{\partial
x_i}-\lambda_k\frac{\partial \varphi_k}{\partial x_i}, \;
\dot{x_i}=\frac{\partial \hat H_0}{\partial
p_i}+\lambda_k\frac{\partial \varphi_k}{\partial p_i}.
\end{equation}
From $\partial \hat H_0/\partial p_i=0$ and
$\partial \varphi_k/\partial p_i=\delta_{ki},$
it follows that the above second equation reduces to
\begin{equation}
\label{Eq:vi}
\dot{x_i}=\lambda_i.
\end{equation}
In this example the Lagrange multiplier $\lambda_i$ is just the
velocity $\dot{x_i}.$

The Poisson brackets
of the constraints 
are
\begin{equation}
\label{Eq:Poisson}
C_{ij}=\{\varphi_i,\varphi_j\}_P= G\epsilon_{ij},
\end{equation}
$C_{ij}$ defined in Eq.~(\ref{Eq:Poisson}) are elements of the
constraint matrix $\mathcal{C}.$ Elements of its inverse matrix
$\mathcal{C}^{-1}$ are $(C^{-1})_{ij}=-\epsilon_{ij}/G.$ The
corresponding Dirac brackets of the canonical variables $x_i$ and
$p_j$ can be determined,
\begin{equation}
\label{Eq:D,x-p}
\{x_i,p_j\}_D=\frac{1}{2}\delta_{ij},\;
\{x_1,x_2\}_D=-\frac{1}{G},\;
\{p_1,p_2\}_D= -\frac{G}{4}.
\end{equation}
The Dirac brackets of $\varphi_i$  with any variables $x_i$ and
$p_j$ are zero so that the constraints (\ref{Eq:C1}) are strong
conditions. It can be used to eliminate dependent variables. If we
select $x_1$ and $p_1$ as independent variables, from the
constraints (\ref{Eq:C1}) we obtain
$x_2=-2p_1/ G,\;p_2= Gx_1/2.$
Introducing new canonical variables $x=\sqrt{2}x_1$  and
$p=\sqrt{2}p_1$ we have $\{x,p\}_D=i\hbar$. The corresponding
quantum commutation relation is $[x,p]=i\hbar$.

The reduced system $(\hat H_0,\hat L_0)$ can be solved as follows.
We define, respectively, the following effective mass and
frequency
\begin{equation}
\label{Eq:mu-om-1}
\hat \mu^{\ast}\equiv \frac{G^2}{2K},\;
\hat \omega^{\ast} \equiv \frac{K}{G},
\end{equation}
then the Hamiltonian $\hat H_0$ reduces to
\begin{equation}
\label{Eq:H0-2}
\hat H_0^{\ast}=\frac{1}{2\hat\mu^{\ast}}p^2 +
\frac{1}{2}\hat \mu^{\ast}\hat \omega^{\ast2}x^2+\mathcal{\hat
E}_{k,0}.
\end{equation}
Eq.~(\ref{Eq:H0-2}) shows that the following annihilation and
creation operators can be introduced
\begin{equation}
\label{Eq:A}
\hat A^{\ast}= \sqrt{\frac{\hat \mu^{\ast}\hat
\omega^{\ast}}{2\hbar}}\;x
+i\sqrt{\frac{1}{2\hbar\hat \mu^{\ast}\hat \omega^{\ast}}}\;p,\;
\hat{A^{\ast}}^\dagger=\sqrt{\frac{\hat \mu^{\ast}\hat
\omega^{\ast}}{2\hbar}}\;x
-i\sqrt{\frac{1}{2\hbar\hat \mu^{\ast}\hat \omega^{\ast}}}\;p.
\end{equation}
They satisfies
$[\hat{A^{\ast}},\hat{A^{\ast}}^\dagger]=1.$
The eigenvalues of the number operator
$\hat{N^{\ast}}=\hat{A^{\ast}}^\dagger\hat{A^{\ast}}$
is $n=0, 1,2, \cdots$. The Hamiltonian $\hat H_0^{\ast}$ reads
\begin{equation}
\label{Eq:H0a} 
\hat H_0^{\ast}
=\hbar\hat \omega^{\ast}\left(\hat{A^{\ast}}^\dagger \hat A^{\ast}
+\frac{1}{2}\right)
+\mathcal{\hat E}_{k,0}.
\end{equation}

Similarly, in the limit of $\hat H_k \to \mathcal{\hat E}_{k,0}$
the Chern-Simons term $\hat J_z$ in Eq.~(\ref{Eq:J1}) reduce to
\begin{equation}
\label{Eq:J2} 
\hat J_z^{\ast}=\hbar\mathcal{\hat J}^{\ast}
\left(\hat{A^{\ast}}^\dagger \hat A^{\ast}
+\frac{1}{2}\right),
\mathcal{\hat J}^{\ast}=1-\xi^2\left(\frac{G\theta}{4\hbar}
+\frac{\eta}{G\hbar}\right).
\end{equation}
The eigenvalues of
$\hat J_z^{\ast}$ is
$\mathcal{\hat J}_n^{\ast}=\hbar\mathcal{\hat J}^{\ast}(n+1/2).$
Its lowest one is
\begin{equation}
\label{Eq:E-J} 
\mathcal{\hat J}_0^{\ast}=\frac{1}{2}\hbar\mathcal{\hat J}^{\ast}.
\end{equation}

In the present case of both position-position and
momentum-momentum noncommuting the second equation of
Eq.~(\ref{Eq:hat-x-p}), up to the first order of $\theta$ and
$\eta$, can be rewritten as
$\hat p_{i}=p_{i}+\eta\epsilon_{ij}x_{j}/2\hbar
=p_{i}+qB_{\eta}\epsilon_{ij}x_{j}/2c.$
It indicates that there is an effective intrinsic magnetic field
$B_{\eta}$ originated from spatial noncommutativity,
\begin{equation}
\label{Eq:B-eff} 
B_{\eta}=\frac{c\eta}{q\hbar}.
\end{equation}
Because $B_{\eta}$ should be vanishingly small, only the external
magnetic field $B$ decreasing to a level of closing to $B_{\eta}$,
the small quantum effects of $B_{\eta}$ from spatial
noncommutativity are manifested obviously.

In the following we demonstrate that, because of the effective
intrinsic magnetic field $B_{\eta}$, in a further limiting process
of diminishing the external magnetic field $B$ to zero the
survived system also has non-trivial dynamics. In this limit the
frequency $\omega_P$ reduces to
$\omega_P=\omega_{\rho}.$
Up to the first order of $\theta$ and $\eta$, we have $\xi =1.$
The effective parameters $M, G, \Omega_P$ and $K$ reduce,
respectively, to
$\tilde M,
\tilde G, \tilde \Omega_P$ and $\tilde K$, which are defined by
\begin{eqnarray}
\label{Eq:M-K1}
\tilde M &\equiv& \left(\frac{1}{\mu}+
\frac{\mu\omega_{\rho}^2\theta^2}{4\hbar^2}\right)^{-1}= \mu,\;
\tilde G\hbar \equiv\mu^2\omega_{\rho}^2\theta+\eta
%
\nonumber\\
\tilde \Omega_P^{\;2}&\equiv&\omega_{\rho}^2+
\frac{\eta^2}{4\mu^2\hbar^2}=\omega_{\rho}^2,\quad\quad
\tilde K\equiv\tilde M\tilde \Omega_P^{\;2}-
\frac{\tilde G^{\;2}}{4\tilde M} =\mu\omega_{\rho}^2.
\end{eqnarray}
We define the following effective mass and frequency
\begin{equation}
\label{Eq:tilde-omega}
\tilde \mu\equiv \frac{\tilde G^2}{2\tilde K},\;
\tilde \omega \equiv \frac{\tilde K}{\tilde G},
\end{equation}
and the  annihilation and creation operators
\begin{equation}
\label{Eq:A1}
\tilde A=\sqrt{\frac{\tilde \mu\tilde \omega}{2\hbar}}\;x+
i\sqrt{\frac{1}{2\hbar\tilde \mu\tilde \omega}}\;p,\;
{\tilde A}^\dagger=\sqrt{\frac{\tilde \mu\tilde
\omega}{2\hbar}}\;x-
i\sqrt{\frac{1}{2\hbar\tilde \mu\tilde \omega}}\;p
\end{equation}
$\tilde A$ and $\tilde A^\dagger$ satisfies $[\tilde A,\tilde
A^\dagger]=1.$ The eigenvalues of the number operator $\tilde
N=\tilde A^\dagger \tilde A$ is $\tilde n=0, 1, 2, \cdots$.
Then, up to the first order of $\theta$ and $\eta$, the $\hat
H_0^{\ast}$ and $\hat J_z^{\ast}$ reduce, respectively, to the
following $\tilde H_0$ and $\tilde J_z$:
\begin{equation}
\label{Eq:H02}
\tilde H_0=\hbar\tilde \omega\left(\tilde A^\dagger \tilde
A+\frac{1}{2}\right),
\tilde J_z=\hbar\mathcal{\tilde J}\left(\tilde A^\dagger \tilde
A+\frac{1}{2}\right),
\mathcal{\tilde J}=1-\frac{\eta}{\tilde G\hbar}.
\end{equation}
The eigenvalues of $\tilde H_0$ and $\tilde J_z$ are,
respectively,
\begin{equation}
\label{Eq:En2}
\tilde E_n = \hbar\tilde \omega\left(\tilde n +
\frac{1}{2}\right),\;
\mathcal{\tilde J}_n=\hbar\mathcal{\tilde J}\left(\tilde
n+\frac{1}{2}\right).
\end{equation}
The term $\eta/\tilde G\hbar$ of $\mathcal{\tilde J}$ in
Eq.~(\ref{Eq:H02}) reads
$\eta/\tilde G\hbar=1/[1+\mu^2\omega_{\rho}^2/(\eta/\theta)].$
Where $\eta/\theta$ is a positive finite constant of dimension
$mass^{-2}time^2$.

In the context of non-relativistic quantum mechanics this can be
elucidated from conditions of guaranteeing the deformed bosonic
algebra
in the case of both position-position and momentum-momentum
noncommuting. The general representations of deformed annihilation
and creation operators $\hat a_i$ and $\hat a_i^\dagger$ at the
deformed level are determined by the deformed Heisenberg-Weyl
algebra (\ref{Eq:xp}) and the deformed bosonic algebra
\begin{equation}
\label{Eq:boson}
[\hat a_1,\hat a_1^\dagger]=[\hat a_2,\hat a_2^\dagger]=1,\;
[\hat a_i,\hat a_j]=0.
\end{equation}
They are \cite{JZZ04a}:
\begin{eqnarray}
\label{Eq:hat-a}
\hat a_i=\sqrt{\frac{1}{2\hbar}\sqrt{\frac{\eta}{\theta}}}
\left(\hat x_i + \frac{i}{\sqrt{\eta/\theta}}\hat p_i\right),
\hat a_i^\dagger=\sqrt{\frac{1}{2\hbar}\sqrt{\frac{\eta}{\theta}}}
\left(\hat x_i-\frac{i}{\sqrt{\eta/\theta}}\hat p_i\right).
\end{eqnarray}
In the limits $\theta,\eta\to 0$ and $\eta/\theta$ keeping finite,
the deformed annihilation operator $\hat a_i$ should reduce to the
undeformed $a_i$ in commutative space.
In the context of non-relativistic quantum mechanics the general
representations of undeformed annihilation and creation operators
$a_i$ and $a_i^\dagger$ are determined by the undeformed
Heisenberg-Weyl algebra
%
%
and the undeformed bosonic algebra
$[a_i,a_j^\dagger]=\delta_{ij},\;
[a_i,a_j]=0.$
They read
\begin{equation}
\label{Eq:a} 
a_i=\sqrt{\frac{1}{2c\hbar}}\left (x_i +ic p_i\right),\;
a_i^\dagger=\sqrt{\frac{1}{2c\hbar}}\left (x_i -ic p_i\right),
\end{equation}
where $c$ is a positive constant. In the limits $\theta,\eta\to 0$
and $\eta/\theta$ keeping finite Eq.~(\ref{Eq:hat-a}) should
reduces to Eq.~(\ref{Eq:a}). It follows that the factor
$\eta/\theta$ in Eq.~(\ref{Eq:hat-a}) should equal $c^{-2}$, that
is,
\begin{equation}
\label{Eq:eta-theta}
\eta=c^{-2}\theta.
\end{equation}
Both noncommutative parameters $\eta$ and $\theta$ should be
extremely small, because modifications to the normal quantum
mechanics originated from spatial noncommutativity should be
vanishingly small. This is guaranteed by Eq.~(\ref{Eq:eta-theta}).

The undeformed Heisenberg-Weyl algebra shows that the equation
$[a_i,a_j]=0$ is automatically satisfied, thus there is not
constraint on the constant $c$. Up till now, how to fix this
constant from fundamental principles is an open issue.

From Eqs.~(\ref{Eq:En2}) and (\ref{Eq:eta-theta}) the dominant
values of the lowest eigenvalue $\mathcal{\tilde J}_0$ and the
interval
$\Delta \mathcal{\tilde J}_n\equiv\mathcal{\tilde J}_{n+1}
-\mathcal{\tilde J}_n $ of $\tilde J_z$
take, respectively,
\begin{equation}
\label{Eq:Jn}
\mathcal{\tilde J}_0=
\frac{\hbar}{2}\frac{1}{1+1/c^2\mu^2\omega_{\rho}^2},\;
\Delta \mathcal{\tilde J}_n=
\hbar\frac{1}{1+1/c^2\mu^2\omega_{\rho}^2}.
\end{equation}
%
Comparing with the corresponding results
$\mathcal{J}_0=\hbar/2$
and
$\Delta \mathcal{J}_n=\hbar/2$
in commutative space, Eq.~(\ref{Eq:Jn}) reveal that in
noncommutative space
$\mathcal{\tilde J}_0<\hbar/2$
and
$\Delta \mathcal{\tilde J}_n<\hbar/2.$
These results are clear signals of spatial noncommutativity.

{\it Towards a Test of Spatial Noncommutativity using Modified
EMS} -- EMS \cite {MW} is used in atomic and molecular physics to
obtain unique information about the motion and correlation of
valence electrons in atoms, molecules and their ions.
EMS involves the measurement of the relative differential cross
section for the (e, 2e) reaction on an atom or molecule as a
function of the electron separation energy and the momenta of
observed electrons. A calculation of the differential cross
section requires a knowledge of the target and ion wave functions.
The reaction can be considered as a measurement of properties of
these wave functions if it is well understood that the cross
section can be calculated within experimental error. The measured
and calculated momentum distributions are different for different
angular momentum states of the electrons in the target, e. g., for
the $s$ state the maximum of the momentum profile appears at the
region of the vanishing momentum, but for the $p$ state the
minimum appears at the same region.

In modified EMS we study the (I, 2I) reaction where $I$ means an
ion. We take the incidental ion as the same type as the target
(trapped) one, and measure the relative differential cross section
of the two outgoing ions as a function of the ion separation
energy and the momenta of observed ions. Here information of the
angular momentum is, different from EMS, for the whole target ion,
not for an electron of the target ion. For our purpose the initial
state of modified EMS is taken as $|i\rangle=|\alpha\rangle
|\chi^{(+)}(\vec k_0)\rangle$, where $|\chi^{(+)}(\vec
k_0)\rangle$ is a distorted wave of the incident ion and
$|\alpha\rangle$ is an angular momentum eigenstate of the target
ion. Information of the angular momentum of the trapped ion is
included in the wave function of the initial state. As a analog of
EMS, the modified EMS (I, 2I) reaction can differentiate between
wave functions of the trapped ions with different angular momenta
from the measured momentum distributions of outgoing ions. For
normal quantum mechanics in commutative space, $\theta=\eta=0$, in
limits of $H_k \to \mathcal{E}_{k,0}$ and subsequent diminishing
the magnetic field $B$ to zero the effective parameter $\tilde
G=0,$ so the effective frequency $\tilde\omega$ and the
annihilation operator $\tilde A$ in Eq.~(\ref{Eq:H02}) cannot be
defined. Using angular momentum wave functions in momentum space
for both normal quantum mechanics and NCQM, calculating the
differential cross sections of (I, 2I) reaction, and comparing
theoretical results with the measured one, we are able to
conclusively determine whether space is noncommutative, i. e., a
positive experimental result shows an evidence of spatial
noncommutativity and a negative one draws a conclusive preclusion
of spatial noncommutativity.

The existing upper bounds of $\theta$ and $\eta$ are,
respectively,
$\theta/(\hbar c)^2\le (10 \;TeV)^{-2}$ \cite{CHKLO}
and
$|\sqrt{\eta}\,|\le 1 \mu eV/c$ \cite {BRACZ}.
From this upper bound of $\eta$ the effective intrinsic magnetic
field
\begin{equation}
\label{Eq:B-eff-1}
B_{\eta}=\frac{c\eta}{q\hbar}\le 10^{-14} T.
\end{equation}
In order to meet the condition of deriving Eqs.~(\ref{Eq:Jn}) the
magnetic field $B$ must be decreased to a level of some orders
less than $B_{\eta}.$
The diminishing level of $B$ is determined by the level of
shielding the background magnetic fields. The field control at
that level seems close to impossible in the foreseeable future.

Recently controlling field at the $10^{-9} T$ level was realized
using magnetic shields of two thick mu-metal cylinders
\cite{SAHTH}. It may relax the field control further to the
challenging, but achievable $10^{-12} T$ level.
It is interesting to consider the case of diminishing the external
magnetic field $B$ to the level of $\sim 10^{-9} T (10^{-12} T).$
Using laser trapping and cooling \cite{SST}, the limit of $\hat
H_k \to \mathcal{\hat E}_{k,0}$ can be reached.
Up to the first order of $\theta$ and $\eta$, the contribution of
the term $\eta/G\hbar$ of $\mathcal{J}^{\ast}$ in
Eq.~(\ref{Eq:J2}) is about
$\eta/G\hbar\sim 10^{-5} (10^{-2}).$
The contribution of the other term $G\theta/4\hbar$ is about
$G\theta/4\hbar\sim 10^{-36} (10^{-39})$
which can be neglected. For normal quantum mechanics in
commutative space, $\theta=\eta=0$, in the limit of $H_k \to
\mathcal{E}_{k,0}$, the lowest angular momentum
$\mathcal{J}_0^{\ast}=\hbar/2.$
The corresponding changes of the lowest angular momentum
$\Delta\mathcal{\hat
J}_0^{\ast}\equiv\mathcal{J}_0^{\ast}-\mathcal{\hat J}_0^{\ast}$
originated from spatial noncommutativity are,
respectively,
\begin{equation}
\label{Eq:J0-J0}
\Delta\mathcal{\hat J}_0^{\ast}
\sim 10^{-5}\hbar (10^{-2}\hbar).
\end{equation}
These results are signals of spatial noncommutativity which are
within limits of the measurable accuracy of the modified EMS.
In this case a positive experimental result shows a primary
evidence of spatial noncommutativity.
One point that should be emphasized is that in this case a
negative one cannot draw a conclusive preclusion of spatial
noncommutativity, but provides an improved upper bound of $\eta$.
%

The limit of the mechanical kinetic energy approaching its lowest
eigenvalue is an important ingredient to obtain the final results.
Now we discuss the consequences if this condition
is not fulfilled. In this case the effects of spatial
noncommutativity are contributed by the second term
$\xi^{2}\theta p_i p_i/2\hbar$
and the third term
$\xi^{2}\eta x_i x_i/2\hbar$
of Eq.~(\ref{Eq:J1}). They can be estimated as follows.
We consider the ion of a mass number in the order $A\sim 10^2$.
Using laser cooling to reduce its average velocity to the order
$\bar v\sim 10^2 ms^{-1}.$  Its average momentum is around the
order
$\bar p\sim 10^{-6} eV/ms^{-1}.$
The average coordinate $\bar x$ can be roughly estimated by the
uncertainty relation:
$\bar x\sim\Delta x$, $\bar p\sim\Delta p$ and
$\bar x\sim\hbar/\bar p.$
Up to the first order of $\theta$ and $\eta$, the contributions of
$\theta\bar p^2/\hbar$
and
$\eta\bar x^2/\hbar\sim\eta\hbar/\bar p^2,$
according to the existing upper bounds of $\theta$ and $\eta$, are
respectively about
$$\theta\bar p^2/\hbar\sim 10^{-20}\hbar,\;
\eta\bar x^2/\hbar\sim 10^{-17}\hbar.$$
They are extremely small.
Testing contributions of spatial noncommutativity at such level is
almost impossible in the foreseeable future.

Technical difficulties involved in the modified EMS (I, 2I)
reaction are as follows:

(i) The differential reaction cross section for the ion-ion
scattering is very small. For a rough estimation using a hard
sphere model with radius $10^{-10} m $, the total cross sections
is about $10^{-20} m^2 $ which depends on the ion's type and
energy \cite{Leco}.

(ii) The efficiency of the coincidence measurements of two
out-going ions is quite small. The efficiency of measuring one
out-going ions is determined by the geometry of the spectrometer
and the open solid angles of ion's going out from the trap. As a
safe region, the open solid angle is estimated as $\sim 1\%$ of
the full $4\pi$ without deteriorating the performance of the trap
\cite{Jagu}.
Therefore, the efficiency of the coincidence measurements of the
two out-going ions is about $10^{-4}$.

It turns out that, like neutrino experiments, the period of the
modified EMS experiment is long. In order to measure one momentum
spectrum in a year, one coincidence measurement per day which
corresponds to a frequency about $10^{-5} s^{-1}$ is necessary.
For this purpose the number of trapped ions and the incident ion
current are, respectively, about $10^{10}$ and $10^{9} s^{-1}.$ In
such cases influence of ion's electric charge and magnetic fields
of moving ions are large which need to be greatly reduced by
special compensation techniques.

(iii) Ions with low energy are easily influenced by the disordered
background electric and magnetic fields. Thus determinations of
momentum distributions of ions through measuring orbits of
scattered ions may lead to large error which should be controlled
at a reasonable level.

(iv) In order to guarantee that the inner structures of the target
and the incident ions are not changed during scattering a careful
selection of the suitable ion's type and beam energy are
necessary.

A more detailed analysis by means of the knowledge of the
noncommutative wave functions shows that
the total cross sections for the ion-ion scattering keeps the same
order as one of using a hard sphere model.
%
%
It is clarified that detailed calculations do not change the basic
characteristics of the results of the above qualitative analysis.
But a detailed analysis of some experimental technique points
are out of the scope of the
present theoretical paper.

Similar to EMS, the modified EMS is one of the most subtle
processes which provides a variety of information for evaluating
the dynamic mechanism of the (I, 2I) reaction. Though test of
spatial noncommutativity using modified EMS is a challenging
enterprise, unlike experiments near the Planck scale $10^{19}
GeV$,
modified
EMS provides a realizable way for measuring noncommutative quantum
effects on normal quantum scales.
%
It is expected that the experimental realization of the proposal
which will be the first step towards a conclusive test of spatial
noncommutativity.

\vspace{0.4cm}

{\bf Acknowledgements}

JZZ's work is supported by NSFC (No 10575037) and SEDF; LKG's work
is supported by NSFC (No 60490280) and NFRPC (No 2005CB724500);
CGN's work is supported by NSFC (No 10575062). JZZ would like to
thank H. P. Liu and X. Shan for helpful discussions.


\clearpage

\begin{thebibliography}{99}

\bibitem{CDS}
A. Connes, M. R. Douglas, A. Schwarz,
J. High Energy Phys. {\bf 02} (1998) 003.

\bibitem{DH}
M. R. Douglas, C. M. Hull,
J. High Energy Phys. {\bf 02} (1998) 008.

\bibitem{AAS}
F. Ardalan, H. Arfaei, M. M. Sheikh-Jabbari,
J. High Energy Phys. {\bf 02} (1999) 016.

\bibitem{CH99}
S-C. Chu, P-M. Ho,
Nucl. Phys. {\bf B550} (1999) 151.

\bibitem{CH00}
S-C. Chu, P-M. Ho,
Nucl. Phys. {\bf B568} (2000) 447.

\bibitem{Sch}
V. Schomerus,
J. High Energy Phys. {\bf 06} (1999) 030.

\bibitem{SE}
N. Seiberg and E. Witten,
J. High Energy Phys. {\bf 09} (1999) 032.

\bibitem{DN}
M. R. Douglas, N. A. Nekrasov,
Rev. Mod. Phys. {\bf 73} (2001) 977 and references there in.

\bibitem{MMP}
Y.-G. Miao, H. J. W. Muller-Kirsten, D. K. Park,
J. High Energy Phys., {\bf 08} (2003) 038.

\bibitem{CST}
M. Chaichian, M. M. Sheikh-Jabbari, A. Tureanu,
Phys. Rev. Lett. {\bf 86} (2001) 2716.

\bibitem{GLR}
J. Gamboa, M. Loewe, J. C. Rojas,
Phys. Rev. {\bf D64} (2001) 067901.

\bibitem{CHKLO}
S. M. Carroll, J. A. Harvey, V. A. Kostelecky, C. D. Lane, T.
Okamoto, Phys. Rev. Lett. {\bf 87} (2001) 141601.

\bibitem{NP}
V. P. Nair, A. P. Polychronakos,
Phys. Lett. {\bf B505} (2001) 267.

\bibitem{HS}
A. Hatzinikitas, I. Smyrnakis,
J. Math. Phys. {\bf 43} (2002) 113.

\bibitem{SS}
A. Smailagic, E. Spallucci,
Phys. Rev. {\bf D65} (2002) 10770.

\bibitem{HK}
P. M. Ho, H. C. Kao,
Phys. Rev. Lett. {\bf 88} (2002) 151602.

\bibitem{BRACZ}
O. Bertolami, J. G. Rosa, C. M. L. de Arag\~ao, P. Castorina, D.
Zappal\`a, Phys. Rev. {\bf D72} (2005) 025010.

\bibitem{JZZ04a}
Jian-zu Zhang,
Phys. Lett. {\bf B584} (2004) 204;
Phys. Rev. Lett. {\bf 93} (2004) 043002;
Phys. Lett. {\bf B597} (2004) 3623;
Qi-Jun Yin, Jian-zu Zhang,
Phys. Lett. {\bf B 613} (2005) 91;
Jian-Zu Zhang,
Phys. Lett. {\bf B639} (2006) 403;
Phys. Rev. {\bf D74} (2006) 124005;
Int. J. Mod. Phys. {\bf 23} (2008) 1393.

\bibitem{MW}
I. E. McCarthy, E. Weigold, Rep. Prog. Phys. {\bf 54} (1991) 789.

\bibitem{trap}
This effective potential $V_{eff}$ can be realized in a Paul
trap(see, for example,
S. L. Brown, G. Gabriels, Rev. Mod. Phys. {\bf 58} (1986) 233;
H. Dehmelt, Rev. Mod. Phys. {\bf 62} (1990) 525). According to its
electrode structure a Paul trap involves an oscillating axially
symmetric electric potential
$$V(\rho,z,t)= (V_d+V_0cos\Omega
t)(\rho^2-2z^2)/(\rho_0^2+2d^2),$$
where $\rho=(x_1^2+x_2^2)^{1/2}$ and $z(=x_3)$ are cylindrical
coordinates;
%
$V_0$ and $V_d$ are, respectively, the amplitude of the
radio-voltage and the dc voltage applied between the electrodes of
the ring and two end caps;
$2d$ is
the separation of the two end caps, and $2\rho_0$ the diameter of
the ring (generally $\rho_0^2=2d^2$);
$\Omega$ is a large radio-frequency. The dominant effect of the
oscillating potential is to add an oscillating phase factor to the
wave function. Rapidly varying terms of time in Schr\"odinger
equation can be replaced by their average values.
%
In the present example, the trap operates in the field of the Paul
trap and a uniform magnetic field ${\bf B}$ aligned along the
$z$-axis
simultaneously, i. e., it is a combined trap
%
(see, e. g., R. J. Cook, D. G. Shankland, A. L. Wells, Phys. Rev.
{\bf A 31} (1985) 564;\;
%
G. Z. Li, Z. Phys. {\bf D 10} (1988) 451;\;
%
R. Blatt, P. Gill, R. C. Thompson, J. Mod. Opt. {\bf 39} (1992)
139;\;
K. Dholakia et al, Phys. Rev. {\bf A47} (1993) 441 and references
there in).
%
Thus for large $\Omega$
we obtain a time-independent effective potential
$$V_{eff}=
\frac{1}{2}\mu(\omega_{\rho}^2\rho^2+\omega_z^2 x_3^2),$$
where
$\omega_{\rho}^2=q^2 V_0^2/8\mu^2\Omega^2 d^4+qV_d/2\mu d^2$
and
$\omega_z^2=q^2 V_0^2/2\mu^2\Omega^2 d^4-qV_d/\mu d^2$
are the macro- frequencies, respectively, in the $(x_1,x_2)$-plane
and $z$ direction.
%
The above effective potential approximation is only valid at high
frequency $\Omega.$
Comparing with the Penning trap, the advantage of the combined
trap is that in the limit of diminishing the magnetic field to
zero it reduces to a Paul trap, thus is still stable as a Paul
trap.



\bibitem{Baxt}
C. Baxter,
Phys. Rev. Lett. {\bf 74} (1995) 514.

\bibitem{JZZ96}
Jian-zu Zhang,
Phys. Rev. Lett. {\bf 77} (1996) 44;
H. J. W. M\"uller-Kirsten, Jian-zu Zhang,
Phys. Lett. {\bf A202} (1995) 241.




\bibitem{SAHTH}
B. E. Sauer, H. T. Ashworth, J. J. Hudson, M. R. Tarbutt, E. A.
Hinds,
%
XX Int. Conf. Atomic Phys. Vol 869, pp. 44-51 (2006),
%
and a
private communication from E. A. Hinds.

\bibitem{SST}
See e. g., F. Shimizu, K. Shimizu, H. Takuma, Opt. Lett. {\bf 16}
(1991) 339. In terms of a number of laser beams and Zeeman tuning
velocities of atoms can reach $1\; m s^{-1}$.

\bibitem{Leco}
For the electron-ion case, see, J. Lecointre et al, J. Phys. {\bf
B41}
(2008) 045201;\;
%
For atom-atom case, see, K. J. Matherson et al, Rev. Sci. Instrum.
{\bf 78} (2007) 073102.

\bibitem{Jagu}
The commercially available ion detectors with a large size can
realize the $~100\%$ detecting efficiency for the out-going ions,
such as the delay line detector, see,
%
O. Jagutzkiet et al,  IEEE Trans. Nucl. Sci., {\bf 49} (2002)
2477.
%
If some thechnique of increasing the outgoing probability of ions
without deteriorating the stable performance of the trap is
realized, the efficiency of the coincidence measurements of the
two out-going ions can be improved.


\end{thebibliography}
\end{document}